\documentclass[AMA,STIX1COL]{WileyNJD-v2}
\setcitestyle{square}

\articletype{Original Article}%

\received{<day> <Month>, <year>}
\revised{<day> <Month>, <year>}
\accepted{<day> <Month>, <year>}

\usepackage{moreverb}

\usepackage{graphicx}
\usepackage{amsmath}
\usepackage{bm}

\usepackage{fancyhdr}
\usepackage{url}
\usepackage{xcolor}
\usepackage{comment}

\graphicspath{ {./images/} }

\begin{document}

\title{Consensus-Based Decentralized Energy Trading for Distributed Energy Resources}

\author[1]{Zhenyu Wang}
\author[2]{Xiaoyu Zhang}
\author[1]{Hao Wang*}

\authormark{Wang et al.}

\address[1]{\orgdiv{Department of Data Science and AI, Faculty of Information Technology}, \orgname{Monash University}, \orgaddress{\state{Melbourne, Victoria}, \country{Australia}}}

\address[2]{\orgdiv{School of Business and Management}, \orgname{Shanghai International Studies University}, \orgaddress{\state{Shanghai}, \country{China}}}

\corres{*Hao Wang, Department of Data Science and AI, Faculty of Information Technology, Monash University, Melbourne, VIC 3800, Australia. \email{hao.wang2@monash.edu}}


\abstract[Abstract]{In smart grids, distributed energy resources (DERs) have penetrated residential zones to provide a new form of electricity supply, mainly from renewable energy. Residential households and commercial buildings with DERs have become prosumers in the local grids, since they can sell surplus power to others. Researches have been initiated to integrate and utilize DERs through better control and communication strategies. With the advances in the Internet of Things (IoT) technology, unprecedented coordination among DERs can be achieved to facilitate energy trading and transactive energy management. However, preventing leakage of users' information during the optimization process keeps challenging researchers, which drives them to develop privacy-preserving energy management systems. In this paper, we develop a fully decentralized transactive energy management using the consensus-based algorithm. To be specific, we design a virtual pool for prosumers to trade energy and exchange information with IoT technologies' support. The consensus-based algorithm enables prosumers to obtain the optimal energy schedule independently in a coordinated manner without revealing any personal data. We use real-world data to perform simulations and validate our developed algorithm. The results show that our consensus-based decentralized transactive energy management strategy is feasible and can significantly reduce the overall system cost.}

\jnlcitation{
\cname{\author{Z. Wang}, \author{X. Zhang} and \author{H. Wang}}
(\cyear{2021}), 
\ctitle{Consensus-Based Decentralized Energy Trading for Distributed Energy Resources}, \cjournal{Energy Convers Econ.}, \cvol{2021;00:1--15}.
}

\maketitle

\section{Introduction}\label{sec1}

The smart grid has been trending across the globe to develop grid intelligence and improve energy efficiency, and distributed energy resources (DERs) have been promoted to deploy at residential homes and commercial buildings to provide new types of sustainable energy sources. DERs, such as solar panels and electric vehicles used locally on the consumption side, transform users into prosumers providing clean energy more efficiently and reliably. Benefited from the evolution of two-way power flow \cite{manusov2018}, prosumers with DERs can save on electricity bills by selling surplus power to the grid or participating in energy trading with other prosumers at a relatively lower price. The emergence of prosumers in the grid also enables a paradigm shift toward transactive energy \cite{yaghmaee2018}. To facilitate transactive energy management of DERs, sufficient information exchange with DERs is often necessary. Due to the nature of low-rate data transmission between DERs, it has been shown that the Internet of Things (IoT) technologies (e.g., Narrowband IoT) can outperform other types of network connection (e.g., WiFi) in DER communication \cite{chen2017narrow}. With the development of IoT technology, the device-to-device (D2D) communications become much more intelligent \cite{Nauman2019}, and DER devices can be monitored and controlled more effectively \cite{choi2017}. This gives rise to lower latency and more considerable scalability in various smart grid applications \cite{li2018smart}. This motivates us to study the following problem in the paper: \textit{what would be a good transactive energy management strategy that promotes efficient utilization of DERs using IoT communications?}

\subsection{Literature Review}

Existing studies have investigated DERs utilization and coordination at the local level and holistic levels. At the local level, DERs with other electric appliances can form an intelligent microgrid or a virtual power plant to improve the grid's energy efficiency and reduce energy cost for prosumers \cite{Vasquez2010}. The energy management system can optimize the energy usage schedule of HVAC and other DERs for each prosumer \cite{fukuta2015}. At the holistic level, transactive energy management can fully utilize renewable energy and other DERs for the whole grid network. The management processes typically require users to share their energy consumption preferences and then schedule cooperatively through a central optimization system. Luna \textit{et al.} designed a centralized system to store and share energy among households, which fully utilizes all DERs in a coordinated manner \cite{Luna2016}. Tenti \textit{et al.} developed a cooperative framework to enhance the grid efficiency by reducing the distribution loss of DERs via central control \cite{Tenti2013}. Yi, Hong and Liu proposed an initialization-free distributed algorithms to eliminate the adaptation work whenever the network structure is changed \cite{yi2016initialization}. However, these centralized management strategies raise privacy concerns, since users' private information would be at risk of leakage when solving a global optimization problem. Particularly for IoT environments, personal data can be easily exploited \cite{Psychoula2018, Eibl2015}, and data leakage may occur \cite{Cha2019}.

Recent studies on distributed transactive energy management have been widely developed to address the challenges of privacy preservation. Parisio \textit{et al.} validated the distributed coordination algorithm under the model predictive control framework, which allowed microgrids to perform local optimization sequentially \cite{Parisio2017}. However, usage information is still disclosed and may be exploited by outside parties. A better way to preserve privacy is to prevent personal data from transmitting to the system. Chang \textit{et al.} employed the contract net protocol and multi-factor methods to carry out the energy exchange optimization alone without including any personal consumption information \cite{Chang2017}. Ge \textit{et al.} designed a novel decomposition algorithm to build a decentralized platform \cite{Ge2020}, where energy transactions among users can be settled to preserve local load information. Le \textit{et al.} adopted a distributed neurodynamic algorithm to optimize the energy trading by exchanging information among neighbors \cite{le2018enabling}. Nizami \textit{et al.} developed a two-stage energy management system that enables energy trading between users under a separated transactive market \cite{Nizami2020}. Still, the success of trading bids heavily depends on users' inclinations, which would bring ineffectiveness in real-world applications.

Researchers have also made efforts on adding protection mechanisms to the information transmission. Aitzhan and Svetinovic \cite{Aitzhan2018} attempted to anonymize the trading transactions using multi-signatures and blockchain technologies. Yang and Wang \cite{yang2020blockchain,yang2021privacy} developed blockchain systems with smart contracts to implement distributed optimization algorithms for transactive energy management of smart homes. Chen \textit{et al.} considered the blockchain technology in economic dispatch problems \cite{chen2021distributed} and peer-to-peer energy trading \cite{chen2021trusted} to secure the reliability without a coordinator. Lu \textit{et al.} adopted Paillier encryption on the private information in the transactive energy system \cite{Lu2020}. Nevertheless, both anonymization and encryption will require extra time and resources to protect user privacy, hindering the deployment of transactive energy systems. 

\subsection{Research Gap and Motivation}

From the existing studies, we can see that different algorithms and technologies have their strengths and weaknesses to protect user information. Most of the proposed systems include central operators to perform the higher-level optimization process. Yang \textit{et al.} proposed a consensus-based algorithm to solve the economic dispatch problem without any reliance on central controllers \cite{Yang2013}. Zhao \textit{et al.} also adopted the consensus-based algorithm to preserve privacy when obtaining the optimal solutions for smart grid energy management \cite{Zhao2017}. However, both of the studies focused on generation units other than prosumers with DERs.
Unlike the above studies, we aim to optimize the energy scheduling for a group of prosumers using the consensus-based algorithm, which would help promote the installation of DERs in residential houses. Celik \textit{et al.} developed a decentralized energy management system for smart homes by allowing coordinated energy sharing, which isolates the pricing mechanism from users \cite{Celik2018}. Wang \textit{et al.} designed a decentralized mechanism to incentivize users to share energy through evaluating their contributions \cite{Wang2019}. In contrast, our market-like pool allows the trading price to be self-adjusted based on prosumers' trading decisions, reflecting the practice of a transactive market. Thus, the need to improve the management process of a fully decentralized system still exists to gain flexibility and protect user privacy at the same time. Our work aims to leverage the advances in IoT technologies enabling device-to-device communications to design privacy-preserving decentralized energy management algorithms for DERs.

\subsection{Contribution and Paper Organization}

In this paper, we present a consensus-based decentralized transactive energy management system with the support of IoT technologies. The main contributions of this work are summarized as follows.

\begin{enumerate}[1)]
    \item Structural Flexibility: Unlike most existing studies of distributed energy management using a fixed communication topology, we establish a decentralized transactive energy system and a market-like virtual pool under flexible communication topology. Our decentralized algorithm enables prosumers to interact and exchange information to determine the optimal energy trading with their neighbors under an IoT network with different D2D communication structures.

    \item Privacy preservation: We develop a consensus-based decentralized algorithm that enables users to reach an agreement on energy trading price and quantity over multiple time slots in the operational horizon. The equilibrium prices obtained for consecutive periods lead to an optimal solution to the transactive energy management problem. The consenting process only requires prosumers to share their energy trading data instead of the complete usage data, which would preserve users' privacy.
    
    \item Real-world practicability: We validate our designed system and algorithm using real-world renewable energy and load data. The results show that the system cost is reduced by $17.30$\%, and the cost of individual prosumers can be reduced by up to $71.54$\%. The results also show the nearest-neighbor D2D communication structure achieves good convergence with much fewer direct connections among prosumers, compared with the all-connected communication structure, demonstrating the practical value of IoT technology for the future decentralized transactive energy system.
\end{enumerate}

The rest of the paper is organized as follows. Section \ref{sec:model} describes the model of the consensus-based decentralized transactive energy system. Section \ref{sec:formulation} formulates the optimization problems for both standalone and coordinated scenarios of the system. Section \ref{sec:solution} elaborates on the design of the consensus-based decentralized algorithm applied to transactive energy management of prosumers. Section \ref{sec:simulation} evaluates the proposed system with extensive simulations, and Section \ref{sec:conclusion} concludes our work.

\section{SYSTEM MODEL}\label{sec:model}

Our system model consists of a power grid, a group of IoT-aided smart homes as prosumers with DERs, and a transactive market that connects prosumers, as shown in Figure \ref{figure1}. We assume that there are $N$ prosumers in the model, which can be denoted as the set $\mathcal{N}=\left\{1,2,\ldots,N\right\}$. We further assume that the system will operate in multiple time slots, which can be denoted as $\mathcal{H}=\left\{1,2,\ldots,H\right\}$, where $H$ is the total number of operating window slots. To be aligned with the market practice, we consider an hour as the operating slot and $H=24$ for each day.

\subsection{Overall Framework}

As shown in Figure \ref{figure1}, we consider that each prosumer is installed with DERs to generate renewable energy (e.g., solar panel or wind turbine) and other electric appliances to meet living requirements (e.g., air conditioner, washing machine, and lights). This is demonstrated by the different icons of devices surrounding the house icon. Each home is also equipped with an advanced smart meter and energy management system to monitor and schedule its own energy use of different types of flexible load. We model it as the \textit{Internal Energy Scheduling Layer}. While our energy management system can optimize the energy schedules for prosumers with different preferences, it can manage the energy trading among prosumers as well. The power exchange  represented by the green ellipse, which supports prosumers to trade energy with each other, forms a virtual pool for sharing energy generated from DERs. The communication required in the pool (e.g., information exchange and transactive activity) can also be facilitated by the IoT technologies applied on the smart meters, which together will constitute an \textit{External Energy Trading Layer}. With transactive energy management, DERs can be fully utilized in the designated area in our model. 

\begin{center}
    \includegraphics[width=0.7 \linewidth]{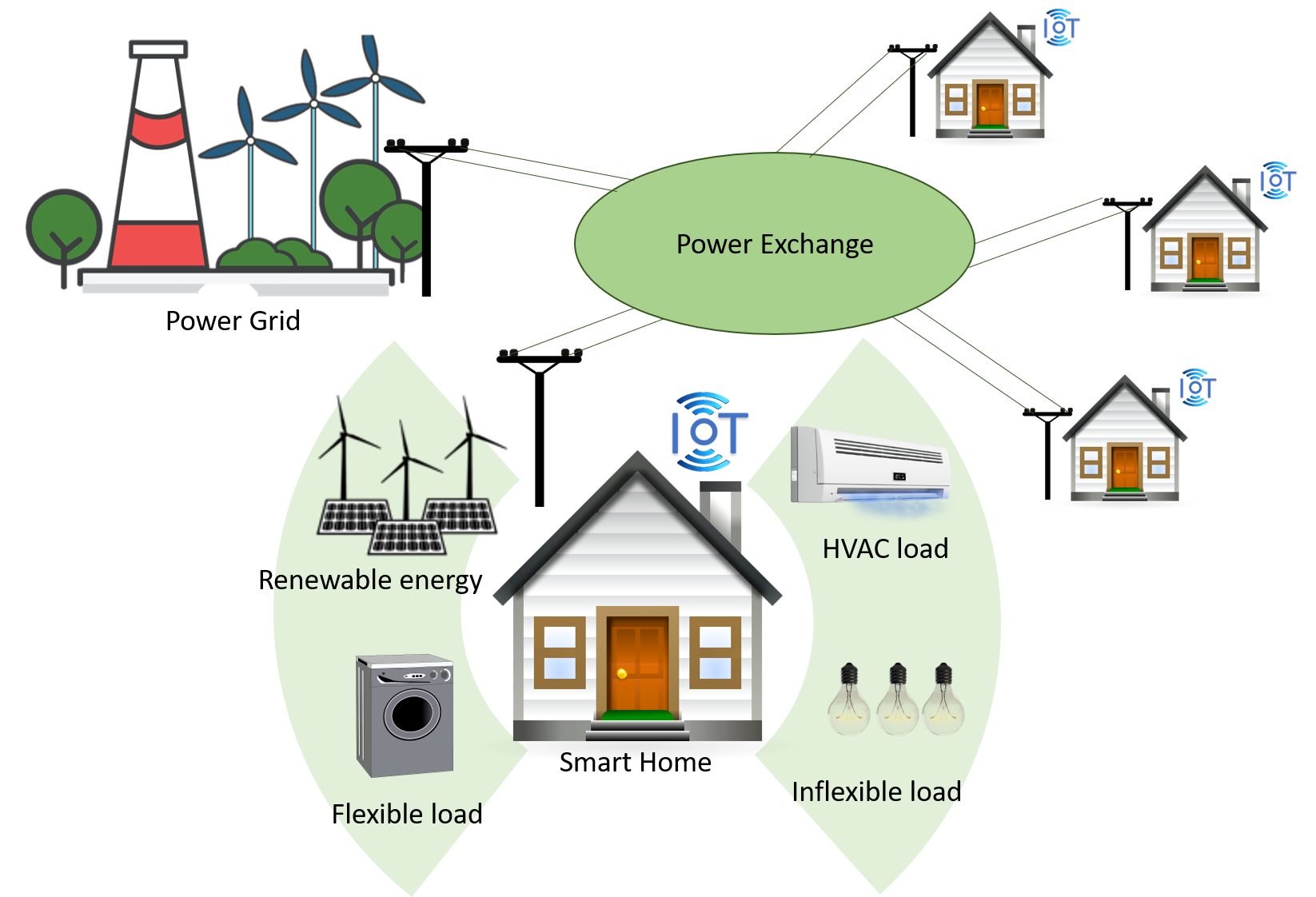}
    \captionof{figure}{The transactive energy management system.}\label{figure1}
\end{center}

\subsection{Internal Energy Scheduling Layer}
We use prosumer and user interchangeably in this paper. As we mentioned earlier, each user's house is equipped with DERs and other appliances, which can be split into two parts: the supply model and the load model. The first part refers to the electricity supply either purchased from the grid or generated from the local DERs. The second part refers to the electricity demand from different appliances in the system.

The supply model for each user $i$ in $\mathcal{N}$ contains two terms $p_{i,t}^\text{G}$ and $p_{i,t}^{\text{RE}}$ in each time slot $t$ in $\mathcal{H}$, where $p_{i,t}^\text{G}$ denotes the amount of energy purchased from the grid, and $p_{i,t}^{\text{RE}}$ denotes the amount of renewable energy generated from DERs. Further, we define $P_i^\text{G}$ to be the maximum capacity to draw electricity from the grid for user $i$ in any time slot, and $P_{i,t}^{\text{RE}}$ to be the available renewable energy from all deployed DERs in user $i$’s home at time $t$. Hence, the two power supply terms $p_{i,t}^\text{G}$ and $p_{i,t}^{\text{RE}}$ should be upper-bounded by $P_i^\text{G}$ and $P_{i,t}^{\text{RE}}$ respectively and should always be non-negative, which can be expressed by the following two constraints:
\begin{align}
    0\le p_{i,t}^\text{G}\le P_i^\text{G},\forall\ i\in\mathcal{N},t\in\mathcal{H}, \label{constraint-grid}
\end{align}
\begin{align}
    0\le\ p_{i,t}^{\text{RE}}\le\ P_{i,t}^{\text{RE}},\ \forall\ i\in\mathcal{N},t\in\mathcal{H}. \label{constraint-renewable}
\end{align}

To promote energy saving and consumption fairness, the grid applies a tiered tariff scheme for users. For example, Austin Energy \cite{austinenergy} has established a five-tier rate structure to charge higher-load consumers at higher rates, as illustrated in Figure \ref{figure2}. We apply the linear approximation to the tiered prices in order to build up our model in more generalized cases, where the unit electricity price is assumed to increase with the load amount linearly. The unit energy price is formulated as
\begin{equation}
    \pi^\text{G}=a^\text{G} p_{i,t}^\text{G}+b^\text{G}, \label{approx-unitPirce}
\end{equation}
where the gradient term $a^\text{G}$ represents the price increase for each unit increase in energy load, and the constant term $b^\text{G}$ represents the base price. Hence, user $i$'s grid cost becomes
\begin{equation}
    C_i^\text{G}=\sum_{t\in\mathcal{H}}{\pi^\text{G}p_{i,t}^\text{G}}=\sum_{t\in\mathcal{H}}\left[a^\text{G}\left(p_{i,t}^\text{G}\right)^2+b^\text{G}p_{i,t}^\text{G}\right]. \label{objective-grid2}
\end{equation}

\begin{center}
    \includegraphics[width= 0.7\linewidth]{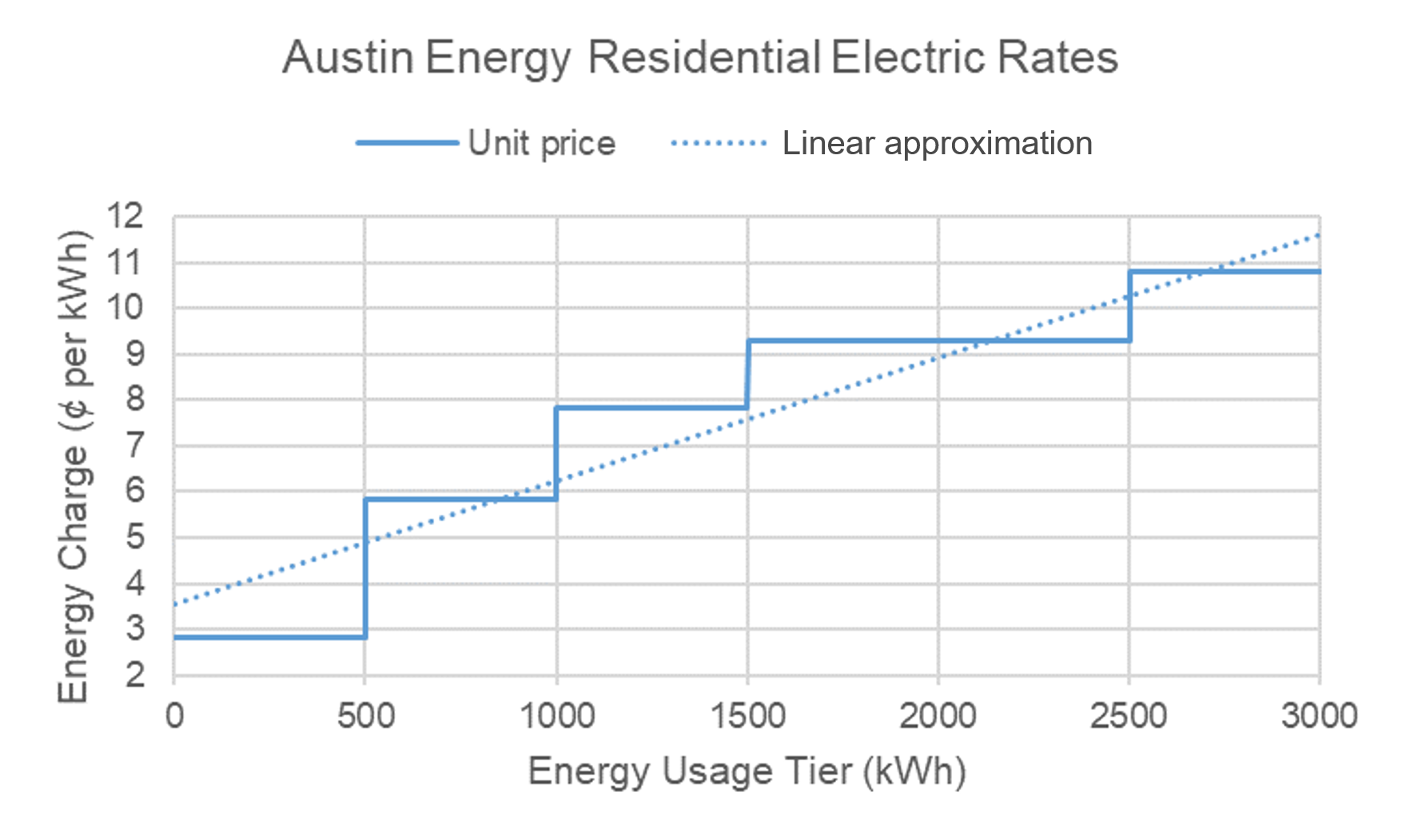}
    \captionof{figure}{An Example for tiered tariff billing with linear approximation (Data source: Austin Energy \cite{austinenergy}).}\label{figure2}
\end{center}

The load model for each user $i$ in time slot $t$ consists of HVAC load, other flexible loads, and inflexible loads, which can be denoted as $p_{i,t}^{\text{AC}}$, $p_{i,t}^\text{F}$ and $p_{i,t}^{\text{IL}}$, respectively. The HVAC load refers to the power consumption of the HVAC unit in adjusting indoor temperature through heating or cooling processes. We define $T_{i,t}^{\text{IN}}$ to be the indoor temperature and $T_{i,t}^{\text{OUT}}$ to be the outdoor temperature for user $i$’s home in time slot $t$. Then the time-series indoor temperature can be formulated according to the thermal dynamics for HVAC units \cite{Cui2019} as
\begin{equation}
    \begin{aligned}
        T_{i,t}^{\text{IN}}=T_{i,t-1}^{\text{IN}}-\frac{1}{\phi_i^C\phi_i^R}\left(T_{i,t-1}^{\text{IN}}-T_{i,t}^{\text{OUT}}+\eta_i\phi_i^Rp_{i,t}^{\text{AC}}\right),\\ \forall\ i\in\mathcal{N},t\in\mathcal{H}, \label{constraint-temperature}
    \end{aligned}
\end{equation}
where $\phi_i^C$ and $\phi_i^R$ denote the operating parameters of the HVAC units, and $\eta_i$ denotes the working mode of the HVAC unit. Cooling gives $\eta_i$ a positive value, and heating gives a negative value. In our simulations, $\phi_i^C$ and $\phi_i^R$ are set to 3.3 and 1.35 respectively for all users. 

HVAC system operates according to the temperature value set by users, which is denoted as $T_i^{\text{REF}}$. When the indoor temperature deviates from the setpoint value, it will cause an uncomfortable feeling to users. We consider such discomfort as the cost in managing HVAC, which is modeled by a quadratic cost function as
\begin{equation}
    C_i^{\text{AC}}=\beta_i^{\text{AC}}\sum_{t\in\mathcal{H}}{\left(T_{i,t}^{\text{IN}}-T_i^{\text{REF}}\right)^2},\ \forall\ i\in\mathcal{N}, \label{objective-hvac}
\end{equation}
where $\beta_i^{\text{AC}}$ denotes the sensitivity coefficient for user $i$.

Also, we define ${\underline{T}}_i^{\text{IN}}$ and ${\overline{T}}_i^{\text{IN}}$ as the lower and upper bounds of user $i$’s tolerable indoor temperature. Hence, the following constraint should be applied to the indoor temperature in time slot $t$:
\begin{align}
    {\underline{T}}_i^{\text{IN}}\le T_{i,t}^{\text{IN}}\le{\overline{T}}_i^{\text{IN}},\ \forall\ i\in\mathcal{N},t\in\mathcal{H}. \label{constraint-hvac}
\end{align}

Other flexible loads, denoted by $p_{i,t}^\text{F}$, refers to the loads that can be shifted over time but cannot be curtailed due to users' living requirements, such as cooking and laundry. We define ${\underline{P}}_{i,t}^\text{F}$ and ${\overline{P}}_{i,t}^\text{F}$ as the lower and upper bounds of user $i$’s flexible load in time slot $t$, and thus the following constraint should be satisfied
\begin{align}
    {\underline{P}}_{i,t}^\text{F}\le p_{i,t}^\text{F}\le{\overline{P}}_{i,t}^\text{F},\ \ \forall\ i\in\mathcal{N},t\in\mathcal{H}. \label{constraint-flexible}
\end{align}

Usually, users have their own most comfortable schedule for flexible activities like cooking or doing laundry. We denote $P_{i,t}^{\text{REF}}$ as the preferred flexible load schedule for user $i$ in time slot $t$. Hence, the total amount of other flexible loads in energy management should always equal to the sum of the preferred schedule, which can be expressed by the following constraint
\begin{equation}
    \sum_{t\in\mathcal{H}}\ p_{i,t}^\text{F}=\sum_{t\in\mathcal{H}}\ P_{i,t}^{\text{REF}},\ \forall\ i\in\mathcal{N}. \label{constraint-flexible2}
\end{equation}

We also consider the user discomfort occurred when the flexible load is shifted away from the preferred schedule. This kind of discomfort represents the cost of managing flexible loads, which can be modeled by the quadratic cost function below
\begin{equation}
    C_i^\text{F}=\beta_i^\text{F}\sum_{t\in\mathcal{H}}{\left(p_{i,t}^\text{F}-P_{i,t}^{\text{REF}}\right)^2},\ \forall i\in\mathcal{N}, \label{objective-flexible}
\end{equation}
where $\beta_i^\text{F}$ denotes the sensitivity coefficient of user $i$ towards the flexible load deviation.

Other inflexible loads $p_{i,t}^{\text{IL}}$ include the electricity consumption of all other appliances that are not adjustable in the system, such as light.

\subsection{External Energy Trading Layer}
In our aforementioned structure, at the same time as scheduling their internal energy, users can trade energy with peers via the virtual power exchange pool using IoT-aided communications. Different from energy-trading studies in \cite{wang2016incentivizing} that forms energy-trading pairs, out work does not require users to identify specific trading partners. Specifically, the transactive energy management system enables users to sell their surplus renewable energy to those having high power demand. The selling parties can have cash inflows to cover some of their electricity bills, whereas the buying parties can purchase energy at a lower price than the regular grid price.

Here, we do not allow prosumers to sell surplus energy to the grid but only to their neighbors. A study on Australian households \cite{Passey2018} has shown that the high penetration of renewable energy at the consumption end will have large financial impacts on centralized power generations but minimal impacts on other consumers who do not have DERs. Therefore, self-consumption is encouraged in a system with a high DER penetration rate to diminish the impacts on the grid network.

We denote the amount of power traded by user $i$ in time slot $t$ as $p_{i,t}^{\text{ET}}$. Here, a positive $p_{i,t}^{\text{ET}}$ represents the status of purchasing power from the others, and a negative $p_{i,t}^{\text{ET}}$ represents the status of selling power to the others. Under the IoT environment, we omit any time lag to process trading information in our model. We also assume that any amount of power exchange can be completed instantly and simultaneously, and there is no electricity loss during the trading, because the geographic distance between any two of the prosumers' houses is negligible. Since the transactive energy cannot be stored for later use in our virtual trading pool, the scheduled trades for all users should be balanced in any time slot $t$, which can be expressed by 
\begin{equation}
    \sum_{i\in\mathcal{N}}\ p_{i,t}^{\text{ET}}=0,\ \forall t\in\mathcal{H}, \label{constraint-trade}
\end{equation}
which means that the sum of energy trading amounts for all users should be cleared by the end of each time slot.

In addition, users should only sell the surplus energy generated from DERs. It will become practically inefficient if the user tries to sell the energy amount over his capacity to generate, because it has to purchase the defective part from the grid at the same time. Therefore, we apply another constraint to prevent users from overselling the renewable energy by lower bounding the energy trading amount $p_{i,t}^{\text{ET}}$.
\begin{equation}
    -P_{i,t}^{\text{RE}}\le\ p_{i,t}^{\text{ET}},\ \forall\ i\in\mathcal{N},t\in\mathcal{H}, \label{constraint-trade2}
\end{equation}
where $-P_{i,t}^{\text{RE}}$ represents the case of selling all the available renewable energy.

The cost/profit of energy trading will be discussed in later sections when formulating the coordinated scenario for our transactive energy management system.

\section{PROBLEM FORMULATION}\label{sec:formulation}
In this section, we consider two scenarios by formulating the optimization problems for both standalone and coordinated energy management of DERs. In the first scenario, each prosumer's energy schedule is optimized independently, where communication and energy trades among users are not required for energy management. In the second scenario, users can trade renewable energy with peers in a coordinated manner to achieve a globally optimal solution for transactive energy management.

\subsection{Standalone Scenario}
In the standalone scenario, the smart meter equipped in each prosumer's house will schedule the internal energy consumption directly without considering any external factors. For each user $i$ in $\mathcal{N}$, power can be supplied either from the grid $p_{i,t}^\text{G}$ or local DERs $p_{i,t}^{\text{RE}}$ to different types of consumption including HVAC load $p_{i,t}^{\text{AC}}$, other flexible load $p_{i,t}^{\text{F}}$ and the inflexible load $p_{i,t}^{\text{IL}}$. To shorten the notation, we redefine the following variables as $\bm{p}_i^{\text{RE}} {=} \{ p_{i,t}^{\text{RE}}, \forall t {\in} \mathcal{H}\}$,
$\bm{p}_i^{\text{G}} {=} \{ p_{i,t}^{\text{G}}, \forall t {\in} \mathcal{H}\}$,
$\bm{p}_i^{\text{AC}} {=} \{ p_{i,t}^{\text{AC}}, \forall t {\in} \mathcal{H}\}$,
$\bm{p}_i^{\text{F}} {=} \{ p_{i,t}^{\text{F}}, \forall t {\in} \mathcal{H}\}$,
and
$\bm{p}_i^{\text{IL}} {=} \{ p_{i,t}^{\text{IL}}, \forall t {\in} \mathcal{H}\}$.

During the energy management process, user $i$'s power supply and demand must be equal in each time slot, which leads to the constraint
\begin{align}
       p_{i,t}^{\text{RE}}+p_{i,t}^\text{G}=p_{i,t}^{\text{AC}}+p_{i,t}^\text{F}+P_{i,t}^{\text{IL}},\ \forall i\in\mathcal{N},t\in\mathcal{H}. \label{constraint-balance}
\end{align}

The left-hand side of \eqref{constraint-balance} represents the total power supply, and the right-hand side of \eqref{constraint-balance} denotes the total demand from the HVAC load $p_{i,t}^{\text{AC}}$, the other flexible load $P_{i,t}^{\text{F}}$ and the inflexible load $P_{i,t}^{\text{IL}}$.

For the standalone energy management, the overall operating cost of user $i$ over the entire time slots $\mathcal{H}$ is
\begin{align}
    C_i^{\text{O}}(\bm{p}_i^{\text{G}}, \bm{p}_i^{\text{AC}}, \bm{p}_i^{\text{F}}) 
     \triangleq
    C_i^{\text{G}}(\bm{p}_i^{\text{G}}) + C_i^{\text{AC}}(\bm{p}_i^{\text{AC}}) + C_i^{\text{F}}(\bm{p}_i^{\text{F}}), \label{objective-operatingcost}
\end{align}
where $C_i^{\text{G}}(\bm{p}_i^{\text{G}})$ denotes the electricity bill from the grid, $C_i^{\text{AC}}(\bm{p}_i^{\text{AC}})$ denotes the user's discomfort cost of HVAC, and $C_i^{\text{F}}(\bm{p}_i^{\text{F}})$ denotes the user's discomfort cost for shifting other flexible load.

The energy management system aims to minimize users' total cost in \eqref{objective-operatingcost}. Therefore, User $i$'s Cost Minimization Problem (\textbf{UCMP}$_i$) for the whole operating period $\mathcal{H}$ can be formulated as follows.

\begin{equation*}
    \begin{aligned}[b]
        & \textbf{UCMP}_i\text{:} \\
        & \min
        && C_i^{\text{O}}(\bm{p}_i^{\text{G}}, \bm{p}_i^{\text{AC}}, \bm{p}_i^{\text{F}}) \\
        & \text{subject to}
        && \text{\eqref{constraint-grid}},\text{\eqref{constraint-renewable}},\text{\eqref{constraint-temperature}},\text{\eqref{constraint-hvac}},\text{\eqref{constraint-flexible}},\text{\eqref{constraint-flexible2}},\text{\eqref{constraint-balance}}\\
        & \text{variables:}
        && \left\{ \bm{p}_i^{\text{RE}}, \bm{p}_i^{\text{G}},  \bm{p}_i^{\text{AC}}, \bm{p}_i^{\text{F}} \right\}.
    \end{aligned}
\end{equation*}

Based on our analysis, user $i$ can locally solve the optimization problem in \textbf{UCMP$_i$} since it is a standard convex optimization. We denote the optimal total cost in problem \textbf{UCMP$_i$} by $\bar{C}_i^{\text{O}}$ and let this value serve as a benchmark cost for the comparison with the costs in the coordinated scenario.

\subsection{Coordinated Scenario}
Unlike the standalone scenario, users not only schedule their internal power supply and demand but also exchange energy externally with other users when needed in the coordinated scenario. Users with surplus renewable energy can sell their energy to other users via the grid; also, users can purchase energy from their neighbors other than the grid to lower the total costs. The energy trading is taken into account, and the load balance constraint becomes
\begin{equation}
    \begin{aligned}
        p_{i,t}^{\text{RE}} {+} p_{i,t}^{\text{G}} {+} p_{i,t}^{\text{ET}} {=} p_{i,t}^{\text{AC}} {+} p_{i,t}^{\text{F}} {+} P_{i,t}^{\text{IL}},  \forall i {\in} \mathcal{N}, t {\in} \mathcal{H}. \label{constraint-balance2}
    \end{aligned}
\end{equation}

Study has shown that the dynamic Lagrangian multiplier can achieve a fast convergence in economic dispatch problems \cite{lai2014decentralized}. We introduce the Lagrange multiplier $\lambda_t$ for the constraints \eqref{constraint-trade}. The Lagrangian function for our optimization problem should be
\begin{equation}
    \begin{aligned}
        \mathcal{L}=
        \sum_{i\in\mathcal{N}}
        \left[
        C_i^{\text{O}}(\bm{p}_i^{\text{G}}, \bm{p}_i^{\text{AC}}, \bm{p}_i^{\text{F}})
        \right]+
        \sum_{t\in\mathcal{H}}
        \left[
        \lambda_{t}
        \sum_{i\in\mathcal{N}}
        p_{i,t}^{\text{ET}}
        \right].
    \end{aligned}
\end{equation}

As we assume that the users can trade energy at any price that they are willing to pay, and such a trading pool can be seen as a perfect market. Hence, the Lagrange multiplier $\lambda_t$ can be considered as the equilibrium price, where the energy trading costs over the entire time slots $\mathcal{H}$ can be defined as:
\begin{equation}
    C_{i}^{\text{ET}} (\boldsymbol{p}_i^{\text{ET}}) = \sum \nolimits_{t \in \mathcal{H}} \lambda_{t} p_{i,t}^{\text{ET}}. \label{objective-tradingpayment}
\end{equation}

In the coordinated energy management system, the system overall cost consists of user's individual operating cost $C_i^{\text{O}}(\bm{p}_i^{\text{G}}, \bm{p}_i^{\text{AC}})$ and the trading cost $C_{i}^{\text{ET}} (\boldsymbol{p}_i^{\text{ET}})$. Therefore, we formulate the Coordinated Cost Minimization Problem (\textbf{CCMP}) for the platform over the whole operating period $\mathcal{H}$ as
\begin{equation*}
    \begin{aligned}[b]
        & \textbf{CCMP}\text{:} \\
        & \min 
        && \sum_{i\in\mathcal{N}} 
        \left[ 
        C_i^{\text{O}}(\bm{p}_i^{\text{G}}, \bm{p}_i^{\text{AC}}, \bm{p}_i^{\text{F}})
        + C_{i}^{\text{ET}} (\bm{p}_i^{\text{ET}}) 
        \right] \\
        & \text{subject to}
        && \text{\eqref{constraint-grid}},\text{\eqref{constraint-renewable}},\text{\eqref{constraint-temperature}},\text{\eqref{constraint-hvac}},\text{\eqref{constraint-flexible}},\text{\eqref{constraint-flexible2}},\text{\eqref{constraint-trade}},\text{\eqref{constraint-trade2}},\text{\eqref{constraint-balance2}}\\
        & \text{variables:}
        &&\{ \bm{p}_i^{\text{RE}}, \bm{p}_i^{\text{G}}, \bm{p}_i^{\text{AC}}, \bm{p}_i^{\text{F}}, \bm{p}_i^{\text{ET}}, ~i \in \mathcal{N} \}.
    \end{aligned} 
\end{equation*}

We see that Problem \textbf{CCMP} jointly solves the optimal internal energy scheduling (of renewable and HVAC load) and the optimal external energy trading for all the users. However, solving \textbf{CCMP} in a centralized manner may cause severe privacy concerns because the users have to reveal all their parameters to the platform. To preserve users' privacy and enable coordinated energy management, we aim to solve Problem \textbf{CCMP} in a decentralized manner in Section \ref{sec:solution}.

\section{DECENTRALIZED ALGORITHM DESIGN}\label{sec:solution}
Consensus-based algorithms can achieve decentralized coordination through finding a certain variable value that is agreed among agents or players \cite{Yang2013}. Since the Problem \textbf{CCMP} is to find the global optimization for all residential users, consensus-based algorithms can be applied to avoid a central coordinator along with its higher-level calculation and its vulnerability in preventing data leakage. As most of the relevant papers have adopted the electricity prices as the consensus variable \cite{Zhao2017, Li2019}, we can use the trading prices among users to achieve consensus in our Problem \textbf{CCMP}. We assume that the energy trading market among users is perfectly competitive, which means the trading prices at any single point of time should be public and at equilibrium. We assume rational prosumers who aim to minimize the expected energy costs under a perfect market, as we aim to automate the optimization process for the energy trading without personal interventions. Since all optimizations will be conducted independently by each user, they do not need to share their personal consumption data with others. The only information to be exchanged among neighbors is the estimated trading price and the estimated energy mismatch, which represents the imbalance in the trading pool.

First of all, we initialize each user by setting up two exchanging parameters, $\lambda_{i,t}\left[0\right]$ and $e_{i,t}\left[0\right]$, which denote the trading price and energy mismatch in time slot $t$ estimated by user $i$ respectively. The initial values can be chosen arbitrarily, because these two parameters are adjusted along the iteration process.

Also, we further set up a communication matrix $W$, where each element $w_{i,j}$ represents the weight that user $i$ put on user $j$'s estimation. Thus, different structures of the communication network can be represented by different matrices. Taking examples of a neighborhood with $5$ users, we illustrate three types of communication network structures represented by $W_1$, $W_2$, and $W_3$, as shown in Figure \ref{figure3}. Type 1 refers to a one-to-many structure where only one user is connected with all the other users. Type 2 refers to a nearest-two structure where all users are only connected with their nearest two neighbors. Type 3 refers to the all-linked structure where all users are connected with each other.

\begin{center}
    \includegraphics[width=0.95 \linewidth]{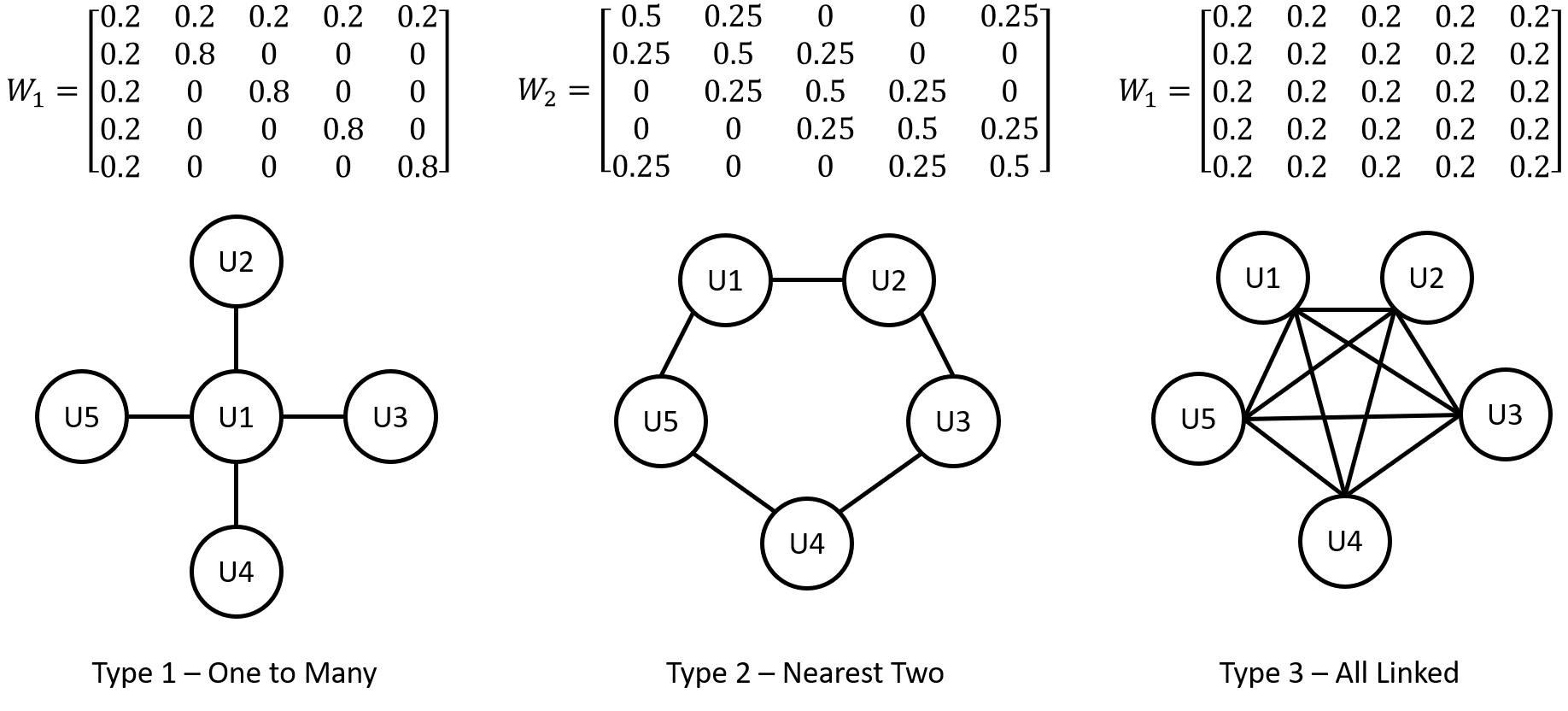}
    \captionof{figure}{Three types of communication network structures with sample communication matrices.}\label{figure3}
\end{center}

Then for each iteration $r$, each user updates its trading price estimates based on its neighbours’ estimates according to the communication matrix $W$, where each element $w_{i,j}$ represents the weight that user $i$ put on user $j$'s estimation. They will adjust this weighted average estimation by a correction term that equals to its energy mismatch estimates multiplied by a learning rate constant $\epsilon$. It means if the energy demand in the trading pool is higher than the energy supply (i.e., the energy mismatch is positive), the trading price should go up, and vice versa, according to the microeconomic principles. Hence, this price update step can be formulated as
\begin{equation}
     \lambda_{i,t}\left[r\right]=\sum_{j\in\mathcal{N}}{w_{i,j}\lambda_{j,t}\left[r-1\right]}+\epsilon e_{i,t}\left[r-1\right], \forall i\in\mathcal{N},t\in\mathcal{H}. \label{update-price}
\end{equation}

With the new estimated trading price $\lambda_{i,t}\left[r\right]$, each user can determine its energy trading schedule by independently finding the optimal solution of its own Cost Minimization Problem, which refers to the individual part of the Problem \textbf{CCMP}, i.e., the overall costs plus the energy trading costs. Specifically, the individual-level problem (\textbf{ILP}$_i$) for the whole operating period $\mathcal{H}$ is formulated as 
\begin{equation*}
    \begin{aligned}[b]
        & \textbf{ILP}_i\text{:} \\
        & \min 
        && C_i^{\text{O}}(\bm{p}_i^{\text{G}}, \bm{p}_i^{\text{AC}}, \bm{p}_i^{\text{F}})
        + C_{i}^{\text{ET}} (\bm{p}_i^{\text{ET}}) \\
        & \text{subject to}
        && \text{\eqref{constraint-grid}},\text{\eqref{constraint-renewable}},\text{\eqref{constraint-temperature}},\text{\eqref{constraint-hvac}},\text{\eqref{constraint-flexible}},\text{\eqref{constraint-flexible2}},\text{\eqref{constraint-trade2}},\text{\eqref{constraint-balance2}}\\
        & \text{variables:}
        &&\{ \bm{p}_i^{\text{RE}}, \bm{p}_i^{\text{G}}, \bm{p}_i^{\text{AC}}, \bm{p}_i^{\text{F}}, \bm{p}_i^{\text{ET}} \}.
    \end{aligned} 
\end{equation*}

After solving \textbf{ILP}, user $i$ will obtain its energy trading decision $\bm{p}_i^{\text{ET}}$. Each user will then update its energy mismatch estimates based on its neighbours’ estimates according to $W$ and a correction term, which is equal to the difference between current and previous energy trading schedules. The energy mismatch update step can be formulated as 
\begin{equation}
    e_{i,t}\left[r\right]=\sum_{j\in\mathcal{N}}{w_{i,j}e_{j,t}\left[r-1\right]}+p_{i,t}^{ET}\left[r\right]-p_{i,t}^{ET}\left[r-1\right], \forall i\in\mathcal{N},t\in\mathcal{H}. \label{update-mismatch}
\end{equation}

The above process will iteratively update users' estimations and optimizations until the consensus is reached. Theoretically, this consensus algorithm's convergence is guaranteed as long as there is a valid trading price corresponding to the minimized cost function \cite{Zhang2012}. Hence, the iteration loop will be terminated when the two most recent iterations' trading prices have a tolerable difference, and the energy mismatches are tolerably small for all users, e.g.,
\begin{equation}
    \left\Vert
    \lambda_i[r] - \lambda_i[r-1]
    \right\Vert_2
    \leq \varepsilon^{\lambda}
    \text{AND}
    \left\Vert
    e_i[r]
    \right\Vert_2
    \leq \varepsilon^{e},
    \forall i \in \mathcal{N}. \label{constraint-iteration}
\end{equation}

Also, Liu \textit{et al.}'s work has proved that the converged results of consensus variable is the unique and optimal solution to the original problem \cite{Liu2020}. Therefore, when all the users reach an agreement for the trading price according to \eqref{constraint-iteration}, their energy schedules will converge to the optimal solution of Problem \textbf{CCMP}. The flow chart shown in Figure \ref{figure4} illustrates the process of our designed consensus-based algorithm.

\begin{center}
    \includegraphics[width=0.95 \linewidth]{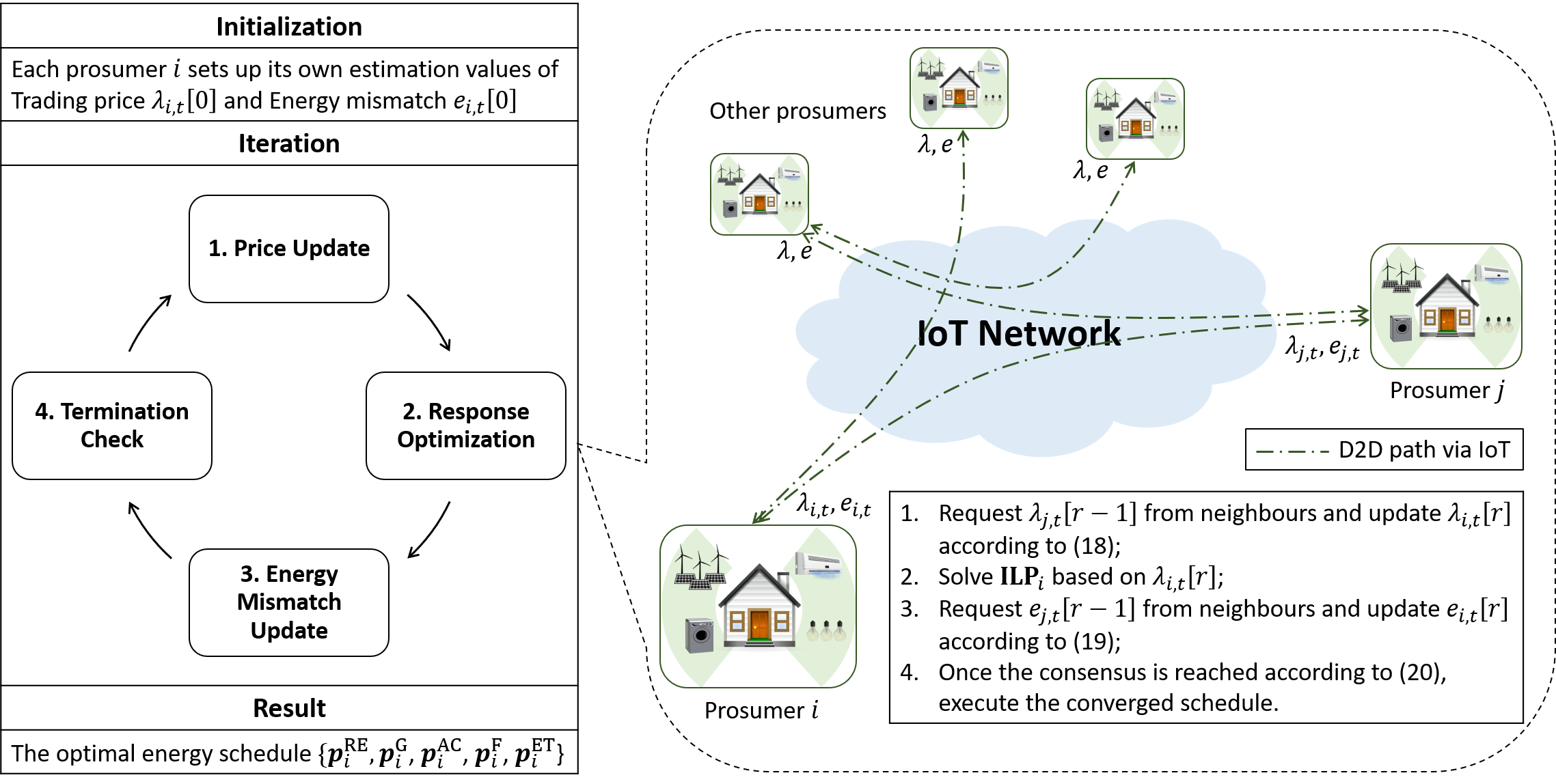}
    \captionof{figure}{The consensus-based algorithm for our transactive energy management system.}\label{figure4}
\end{center}

\section{Simulation Results and Discussions}\label{sec:simulation}
In this section, we evaluate the transactive energy management system under both standalone and coordinated scenarios through extensive numerical simulations with real data of renewable generation \cite{wang2014hybrid,wang2015joint} and load from homes \cite{pecan}. The simulation data include users’ daily electricity usage, renewable energy generation, and outdoor temperature.
To validate the consensus-based decentralized optimization algorithm developed in Section 4, we simulate a group of 10 users, and their corresponding smart homes are equipped with DERs and other appliances according to the system model we set up in Section~\ref{sec:model}.

\subsection{Algorithm Convergence}
We firstly evaluate the performance of the decentralized optimization algorithm in Section 4 with 10 users. The convergence test of the consensus-based decentralized algorithm mainly targets two aspects: \textit{the trading price consensus} and \textit{the trading balance} in the trading pool.

The trading price consensus means that all users will agree to the same energy trading price in any time slot. In the designed algorithm, the convergence of trading price can be observed directly by the changes to users' estimations (i.e., $\lambda_{i,t}$) throughout the iteration process. As shown in Figure \ref{figure5}, we randomly select 6 time slots in the Day 1 simulation and plot the estimated trading price for each user with respect to the number of iterations. We can see that the trading price estimated by each user can always converge to an equilibrium value after around $30$ iterations through IoT-aided communications.

\begin{center}
    \includegraphics[width=0.84\linewidth]{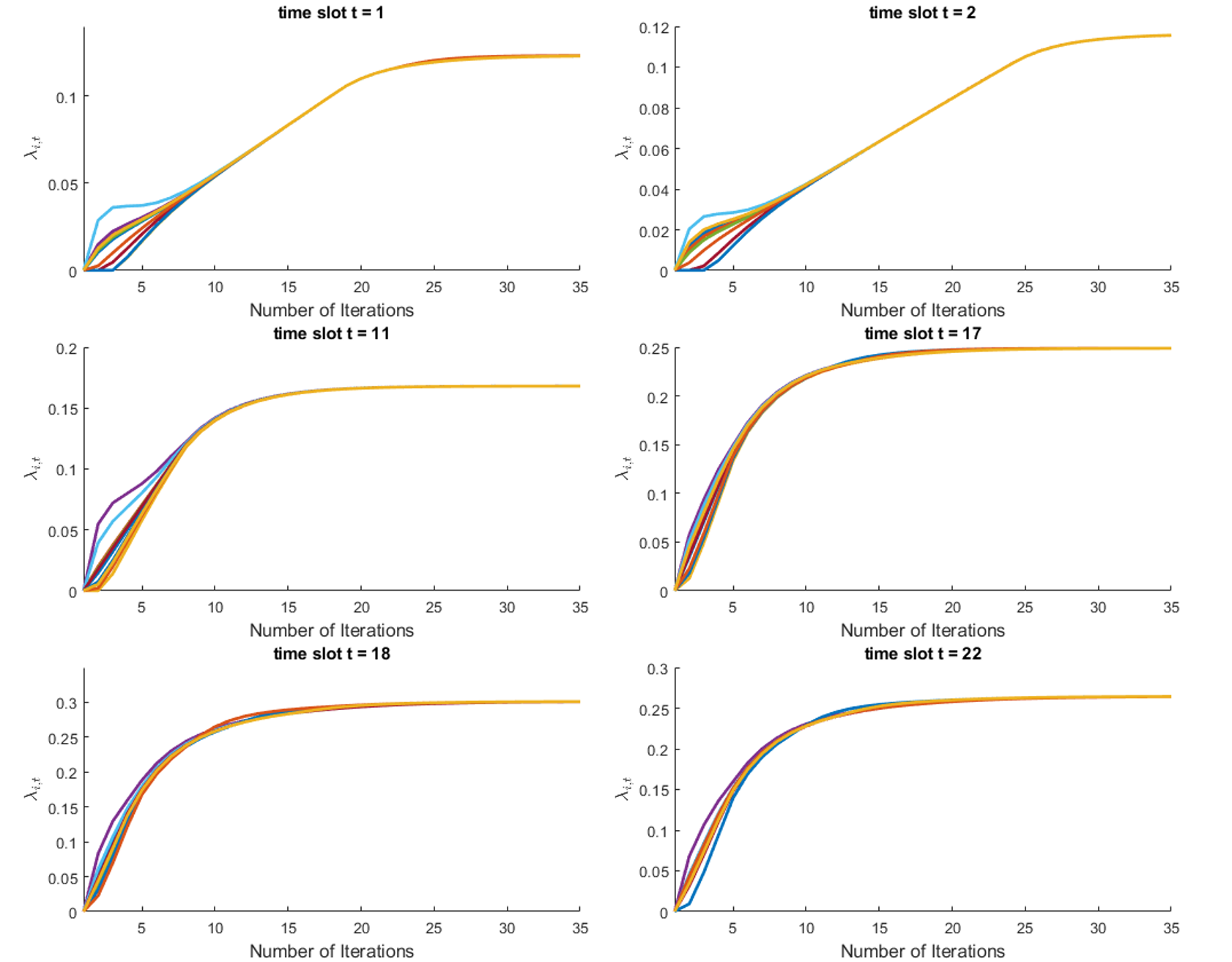}
    \captionof{figure}{The consensus process of trading prices $\lambda_{i,t}$ in selected time slots in the Day 1 simulation.}\label{figure5}
\end{center}

To further evidence the flexibility of our proposed model, we attempted to validate it under different kinds of communication network structures and larger user scales. As shown in Figure \ref{figure6}, we similarly plot the estimated trading price for two selected time slots for three scenarios: (a) 10 users under the one-to-many structure, (b) 10 users under the nearest-four structure, and (c) 50 users under the all-linked structure. We can observe that the convergence still holds under different types of communication networks and different scales of the user group. However, when there are insufficient links between users, it will cost more iterations, e.g., over $150$ times under scenario (a), which is reasonable since coordination works better with more links.

\begin{center}
    \includegraphics[width=1 \linewidth]{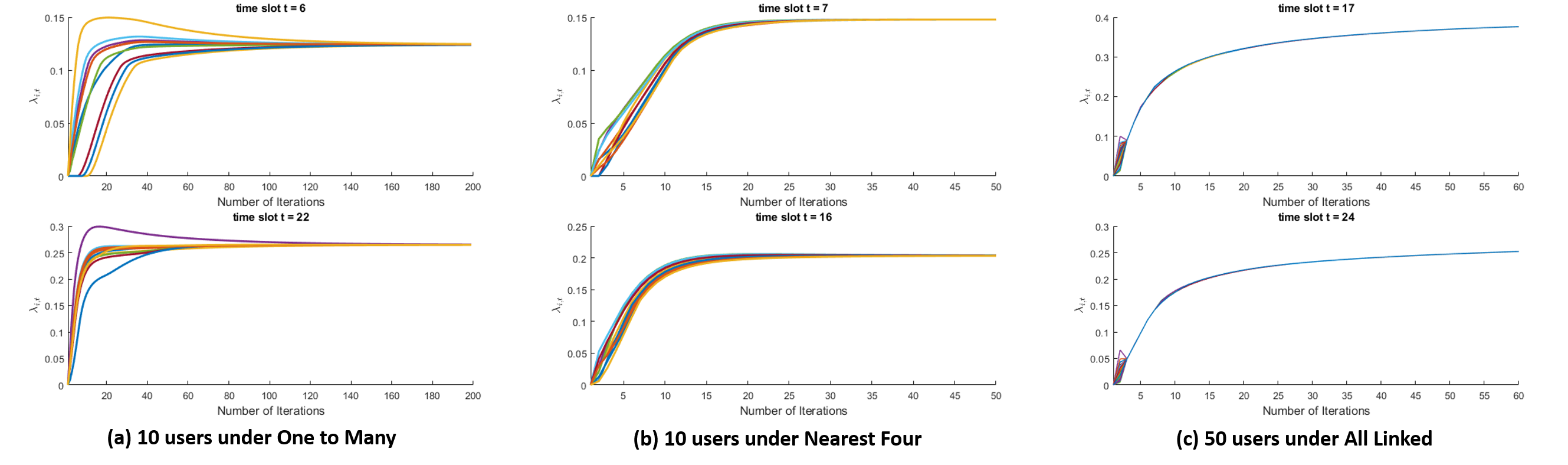}
    \captionof{figure}{The convergence of models with different users and communication networks.}\label{figure6}
\end{center}

The other aspect, trading balance, refers to the supply-and-demand balance of users' optimal energy exchange in any time slot. In microeconomic theories, the total supply amount should be equal to the total demand amount for a commodity at equilibrium under the perfectly competitive market. Hence, the supply energy should be equal to the demand energy within our virtual trading pool when the algorithm converges. This can be tested by calculating the total amount of actual mismatch in the power exchange for all users and in all time slots (i.e., $\sum_{t \in \mathcal{H}} \sum_{i \in \mathcal{N}} p_{i,t}^{\mathrm{ET}}$) throughout the iteration process. As shown in Figure \ref{figure7}, we plot the sum of actual mismatches for all time slots with respect to the number of iterations in the Day 1 simulation under the four scenarios we obtained previously. Figure \ref{figure7}(a) can compare the convergence among three types of communication networks. The one-to-many structure requires the longest time to converge, whereas the all-linked structure is the fastest. Again, it illustrates that more links among the network will help speed up the convergence process given the same number of users. Figure \ref{figure7}(b) shows it takes much more iterations if we have more users in the network, e.g., $50$ users will need about $500$ times of iterations.

\begin{center}
    \includegraphics[width=1 \linewidth]{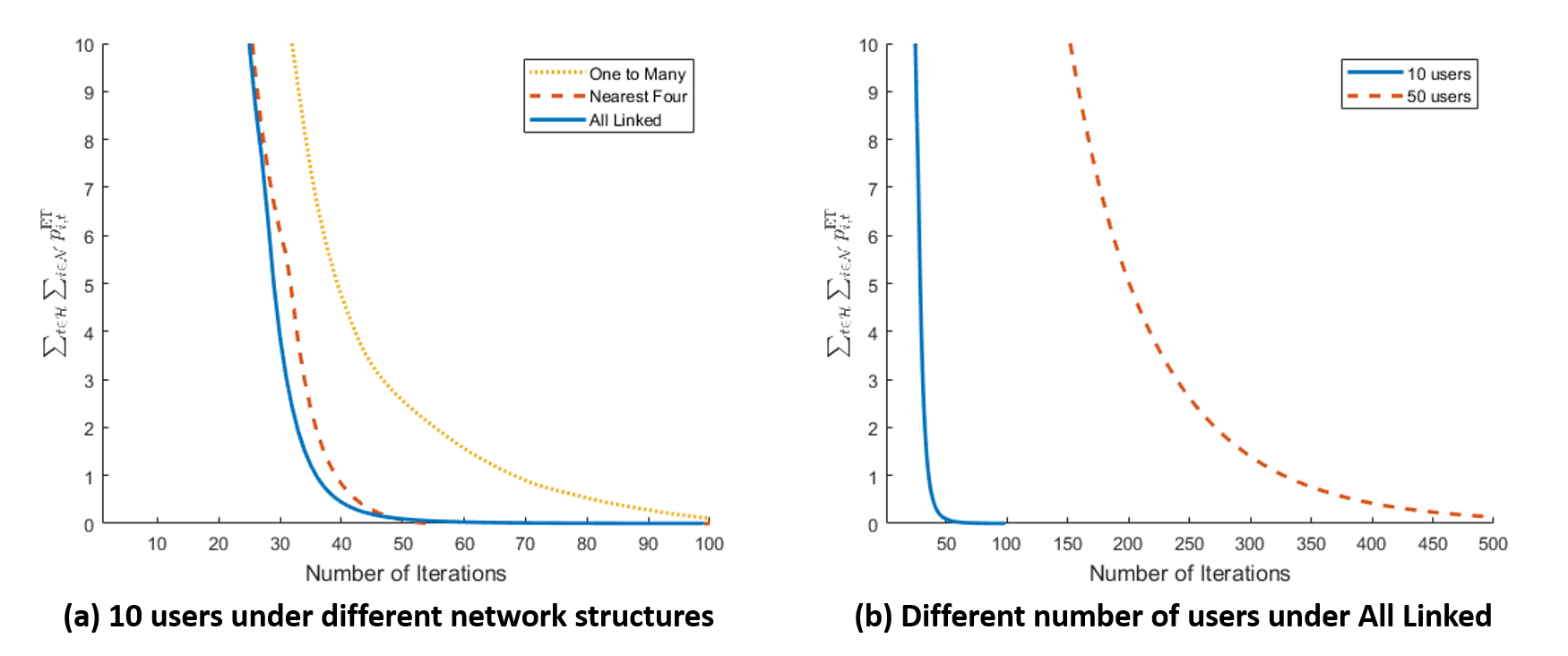}
    \captionof{figure}{The sum of actual mismatch for all time slots in the Day 1 simulation under different networks.}\label{figure7}
\end{center}

Regarding the time for transmitting data through IoT communication, our model only requires prosumers to request two data values (i.e., the estimated trading price and the estimated energy mismatch) from linked neighbors. Hence, the time spent to complete one iteration will relate to the number of links in the networks. In other words, a more complex network will take a longer time to process each iteration. In our simulation, $30$ iterations take around $3$ seconds for $10$ users under an all-linked network.

\subsection{Optimal Energy Management}
Secondly, we illustrate the optimal solution obtained from the model simulation for both internal and external layers under the coordinated scenario.

For \textit{Internal Energy Scheduling Layer}, as shown in Figure \ref{figure8}, we plot the final energy scheduling outputs for all $10$ users when the convergence is achieved. We can see that the Grid Usage, which represents the amount of energy purchased from the grid, is managed to be much stable at most of time for all users. In comparison with Renewable Energy, which indicates the utilization of DERs, the patterns for these two energy sources show a complementary relationship throughout the simulation week. It suggests that the stabilization in the grid is mostly benefited from the optimum use of renewable energy generated from DERs. The plot for HVAC Load has similar peak hours to the Renewable Energy, which demonstrates that the energy management system can adjust the undulation of electricity consumption according to the effectiveness of DER generation.
\begin{center}
    \includegraphics[width=0.8 \linewidth]{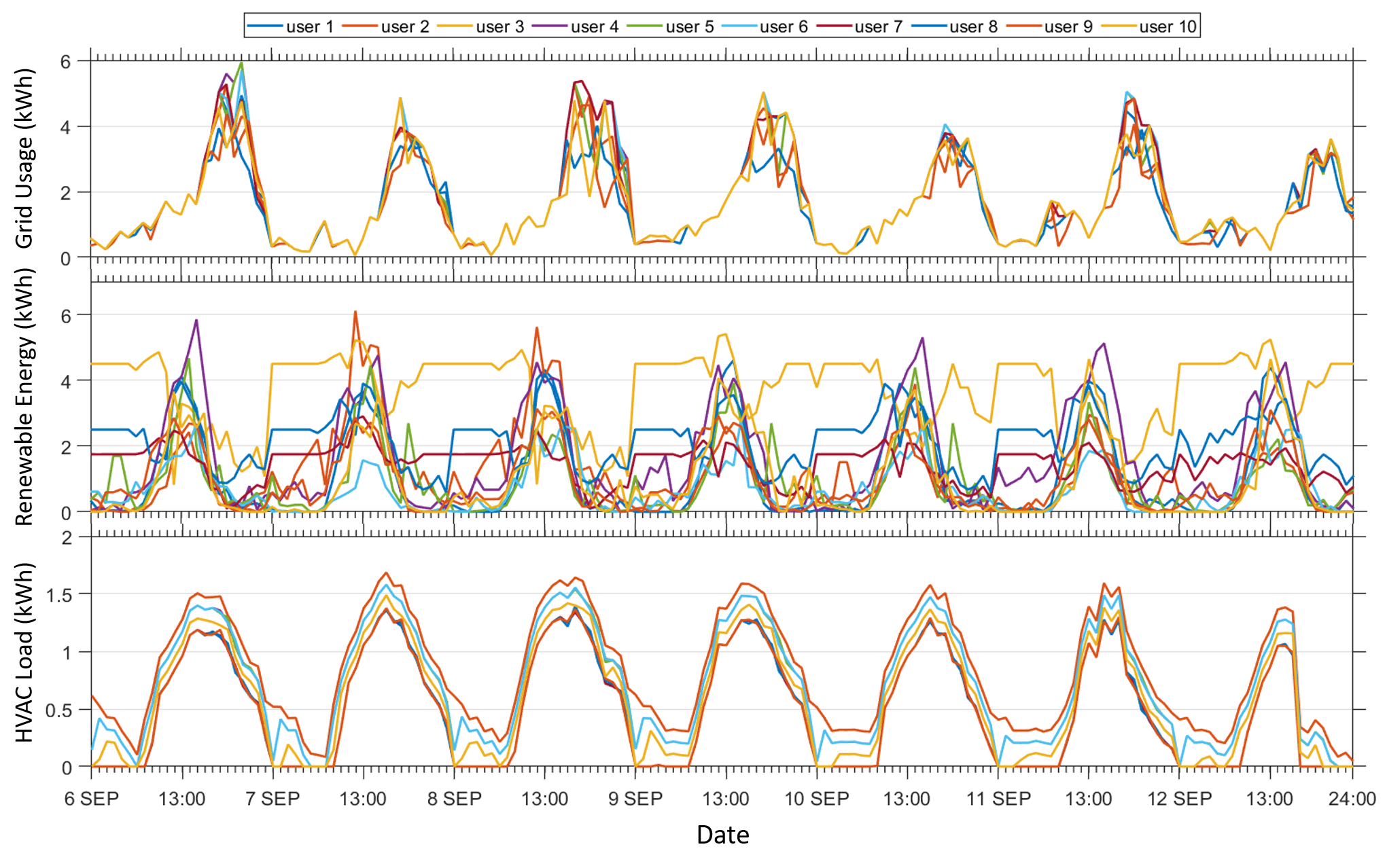}
    \captionof{figure}{The optimal energy schedule under coopertative management in one-week simulation.}\label{figure8}
\end{center}

For \textit{the External Trading Layer}, as shown in Figure \ref{figure9}, we plot the final energy trading arrangements for selected users in the one-week simulation, where the orange bars with positive value represent the buying parties in the trading and the purple bars with negative value represents the selling parties. We can observe that there are $3$ typical types of users in energy trading. The first type refers to those who are buying a lot of energy from other peers most of the time, such as User 4 shown in Figure \ref{figure9}(a). It means that their DERs cannot generate enough renewable energy to cover their own demand. The second type refers to those who mostly behave as energy sellers, such as User 3 shown in Figure \ref{figure9}(b). They often have surplus renewable energy for sale, because they have a high DER capacity to cover their daily load. The rest of the users who have mixed positions as buyers and sellers throughout the scheduling period are classified as the third type like User 10 shown in Figure \ref{figure9}(c). They can exchange power with others as they needed, such that all renewable energy will be fully utilized locally in the neighborhood.

\begin{center}
    \includegraphics[width=0.8 \linewidth]{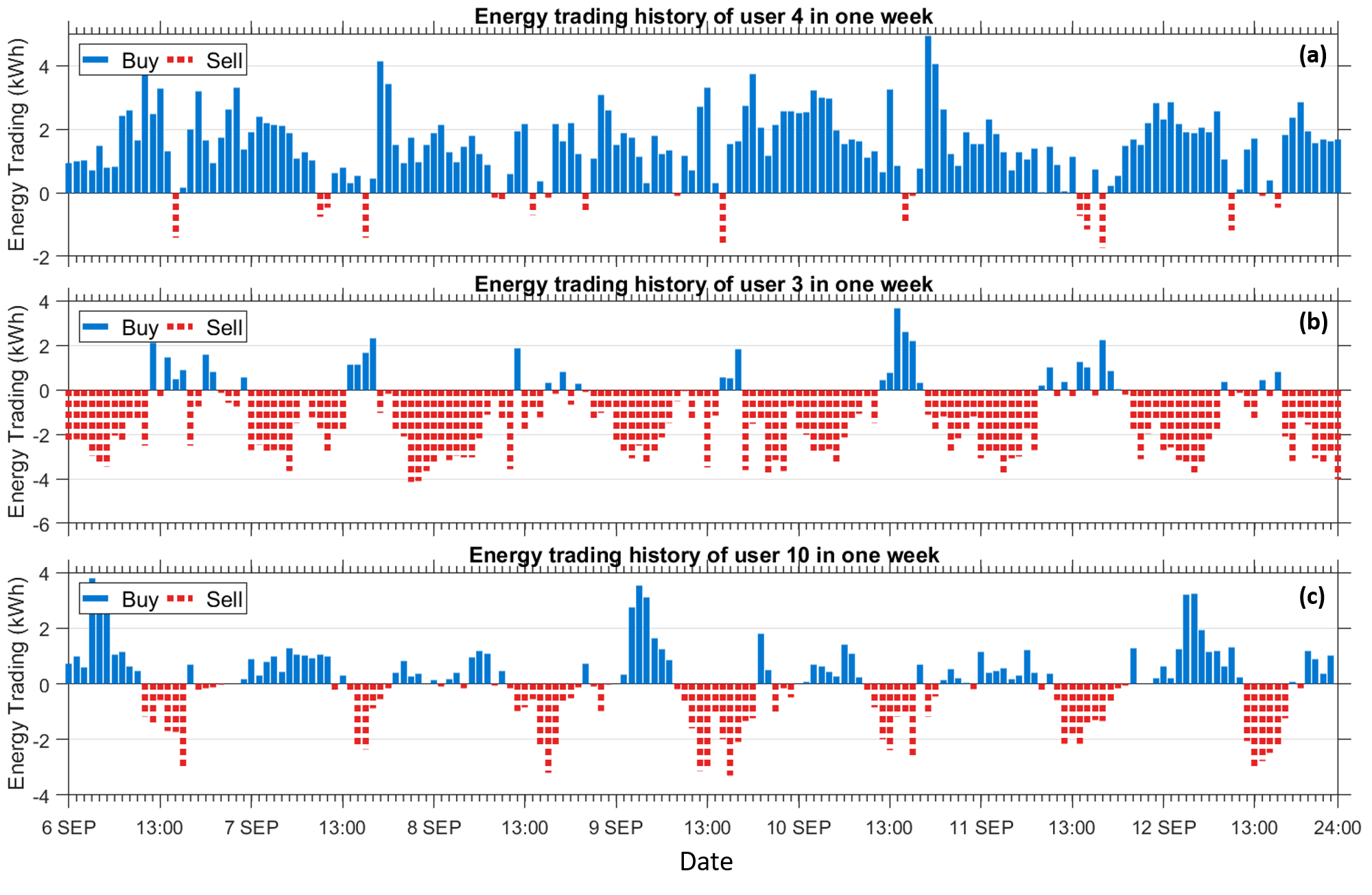}
    \captionof{figure}{The optimal energy trading history for selected users under coopertative management in one-week simulation.}\label{figure9}
\end{center}

\subsection{Energy and Cost Reduction}
Lastly, we evaluate the cost reduction for users after implementing our transactive energy management system by comparing the standalone and coordinated scenarios.

As shown in Figure \ref{figure10}, the blue line indicates the scheduled amount of grid energy purchased by User 6 under standalone management. The red line shows his new schedule after joining the coordinated trading with his peers. We can see that the electricity purchased from outside power plants has been significantly reduced during the entire simulated week, given that the consumptions are not much affected. Hence, the goal of energy-saving can be well achieved under our coordinated strategy.

\begin{center}
    \includegraphics[width=0.8 \linewidth]{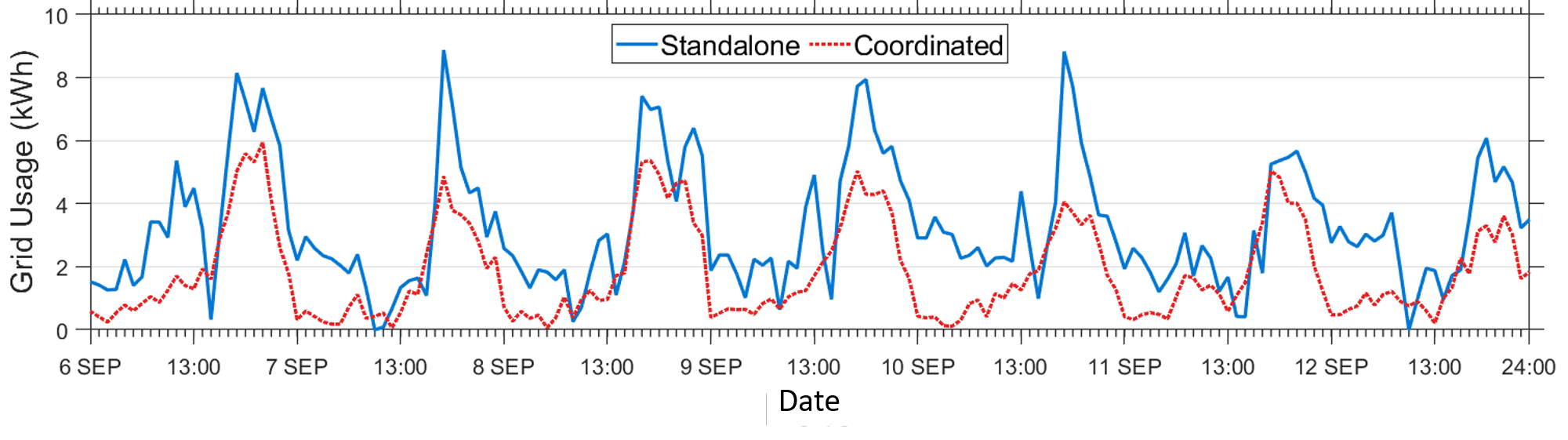}
    \captionof{figure}{The comparison of grid power purchased by User 4 under standalone and coordinated scenarios.}\label{figure10}
\end{center}

As shown in Figure \ref{figure11}, we plot the overall costs for each user under the two scenarios (orange bars for the standalone scenario and purple bars for the coordinated scenario) and compute their percentage changes (the number above bars). The results show that the coordinated management system can lower the overall cost for all participated users. The cost reduction for individual producers can be reduced from 8.00\% to 71.54\% at most, and clearly, users who sell more in energy trade can benefit more. The system's total cost can be reduced by 17.30\%, which demonstrates a remarkable economic advantage for our transactive energy management system.

\begin{center}
    \includegraphics[width=0.8 \linewidth]{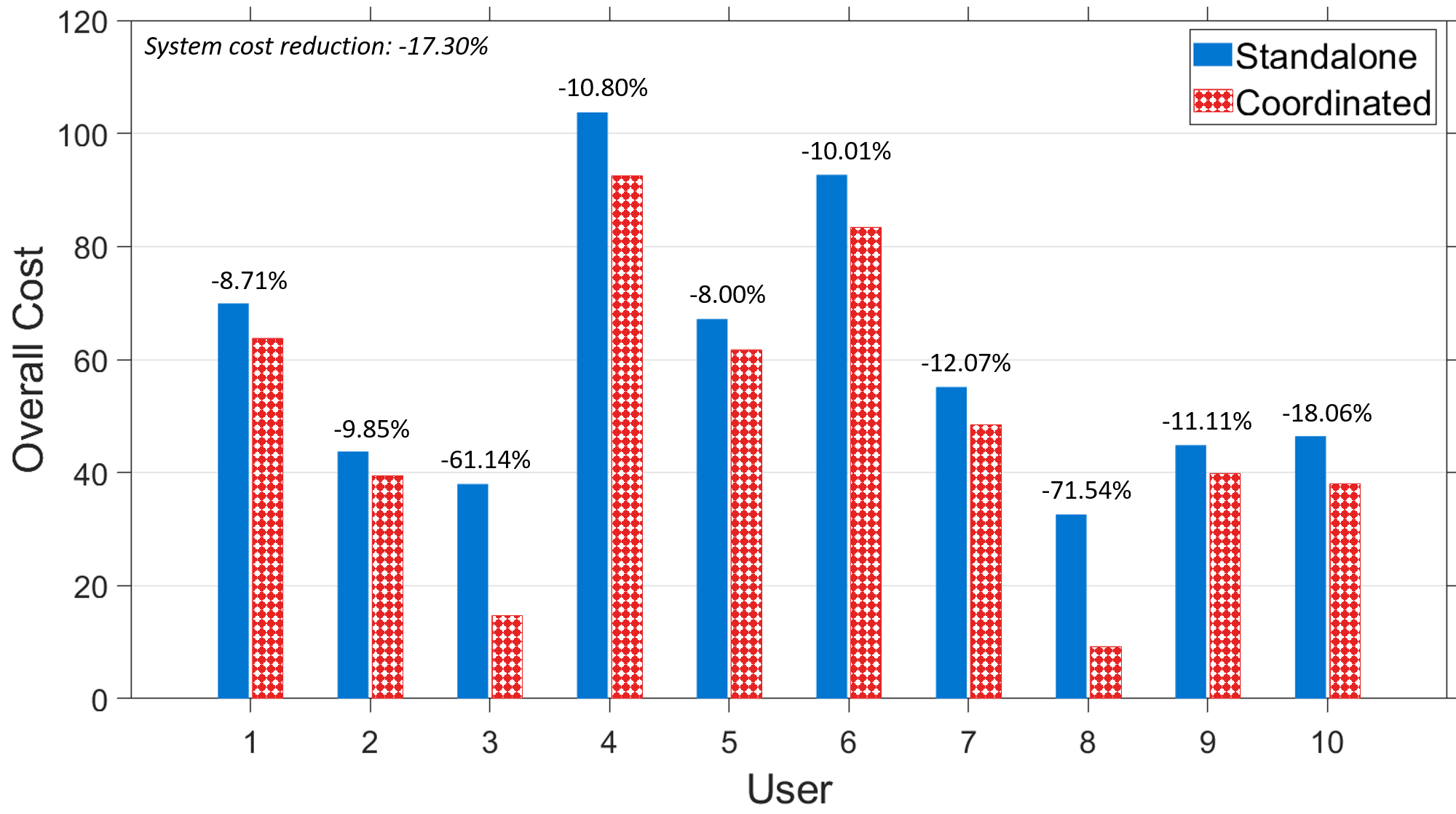}
    \captionof{figure}{The overall costs for all 10 users under standalone and coordinated scenarios. (The percentage number above bars refers to the cost reduction.)} \label{figure11}
\end{center}

\section{Conclusion}\label{sec:conclusion}
This paper presented a consensus-based decentralized transactive energy management strategy that enables prosumers with DERs to trade energy with each other in an automated manner. We adopted a consensus-based algorithm to achieve decentralized coordination and hence to improve the utilization of local DERs.

\begin{itemize}
    {
    \item Our designed algorithm does not require any central coordinator, because it can be completely set up with D2D communication supported by IoT technologies. Also, our model will be maneuverable to different types of D2D communication network structures.
    \item Information transmitted between the IoT-aided smart meters only involves individual estimations on trading prices and energy mismatch during iterations, which means personal usage information will not be shared with others, and users' privacy can be well preserved.
    \item Our simulation results showed that our system is feasible under different network structures and scalable to different sizes of grids. The costs for both the gird and all users can be significantly reduced under our model.
    \item For future work, we will improve the iteration process of the consensus-based algorithm when implementing it in a larger community with more homes. Also, we will attempt to integrate our transactive energy management system with blockchain technology (e.g., smart contract) to further enhance the automation process and secure personal transactive information during energy trading.
    }
\end{itemize}

\bibliographystyle{NJDnatbib}
\bibliography{ref.bib}%

\begin{thebibliography}{10}
\providecommand \doibase [0]{http://dx.doi.org/}%

\bibitem{manusov2018}
{Manusov} VZ, {Khasanzoda} N, {Atabaeva} LS. Two-Way Energy Flow Optimization
  Based on Smart Grid Concept. In:  {\it 2018 International Conference on
  Industrial Engineering, Applications and Manufacturing (ICIEAM), May 15-18};
  2018; Moscow, Russia\string: 1-6

\bibitem{yaghmaee2018}
{Yaghmaee Moghaddam} MH, {Leon-Garcia} A. A fog-based internet of energy
  architecture for transactive energy management systems. {\it IEEE Internet of
  Things Journal} 2018\string; 5(2)\string: 1055-1069.
\newblock \href {\doibase 10.1109/JIOT.2018.2805899} {doi:
  10.1109/JIOT.2018.2805899}

\bibitem{chen2017narrow}
{Chen} M, {Miao} Y, {Hao} Y, {Hwang} K. Narrow Band Internet of Things. {\it
  IEEE Access} 2017\string; 5\string: 20557-20577.
\newblock \href {\doibase 10.1109/ACCESS.2017.2751586} {doi:
  10.1109/ACCESS.2017.2751586}

\bibitem{Nauman2019}
{Nauman} A, {Jamshed} MA, {Ahmad} Y, {Ali} R, {Zikria} YB, {Won Kim} S. An
  Intelligent Deterministic D2D Communication in Narrow-band Internet of
  Things. In:  {\it 2019 15th International Wireless Communications Mobile
  Computing Conference (IWCMC), Jun. 24-28}; 2019; Tangier, Morocco\string:
  2111-2115

\bibitem{choi2017}
{Choi} M, {Lee} I. Distributed energy resources monitoring and control based on
  IoT protocol. In:  {\it 2017 International Conference on Information and
  Communication Technology Convergence (ICTC), Oct. 18-20}; 2017; Jeju Island,
  Korea\string: 1115-1117

\bibitem{li2018smart}
{Li} Y, {Cheng} X, {Cao} Y, {Wang} D, {Yang} L. Smart Choice for the Smart
  Grid: Narrowband Internet of Things (NB-IoT). {\it IEEE Internet of Things
  Journal} 2018\string; 5(3)\string: 1505-1515.
\newblock \href {\doibase 10.1109/JIOT.2017.2781251} {doi:
  10.1109/JIOT.2017.2781251}

\bibitem{Vasquez2010}
{Vasquez} JC, {Guerrero} JM, {Miret} J, {Castilla} M, {de Vicuña} LG.
  Hierarchical Control of Intelligent Microgrids. {\it IEEE Industrial
  Electronics Magazine} 2010\string; 4(4)\string: 23-29.
\newblock \href {\doibase 10.1109/MIE.2010.938720} {doi:
  10.1109/MIE.2010.938720}

\bibitem{fukuta2015}
{Fukuta} M, {Matsui} K, {Ito} M, {Nishi} H. Proposal for home energy management
  system to survey individual thermal comfort range for HVAC control with
  little contribution from users. In:  {\it 2015 IEEE 13th International
  Conference on Industrial Informatics (INDIN), Jul. 22-24}; 2015; Cambridge,
  UK\string: 658-663.

\bibitem{Luna2016}
{Luna} AC, {Diaz} NL, {Graells} M, {Vasquez} JC, {Guerrero} JM. Cooperative
  energy management for a cluster of households prosumers. {\it IEEE
  Transactions on Consumer Electronics} 2016\string; 62(3)\string: 235-242.
\newblock \href {\doibase 10.1109/TCE.2016.7613189} {doi:
  10.1109/TCE.2016.7613189}

\bibitem{Tenti2013}
{Tenti} P, {Costabeber} A, {Caldognetto} T, {Mattavelli} P. Improving microgrid
  performance by cooperative control of distributed energy sources. In:  {\it
  2013 IEEE Energy Conversion Congress and Exposition, Sep. 15-19}; 2013;
  Denver, CO, USA\string: 1647-1654

\bibitem{yi2016initialization}
Yi P, Hong Y, Liu F. Initialization-free distributed algorithms for optimal
  resource allocation with feasibility constraints and application to economic
  dispatch of power systems. {\it Automatica} 2016\string; 74\string: 259--269.

\bibitem{Psychoula2018}
{Psychoula} I, {Singh} D, {Chen} L, {Chen} F, {Holzinger} A, {Ning} H. Users'
  Privacy Concerns in IoT Based Applications. In:  {\it 2018 IEEE SmartWorld,
  Ubiquitous Intelligence Computing, Advanced Trusted Computing, Scalable
  Computing Communications, Cloud Big Data Computing, Internet of People and
  Smart City Innovation, Oct. 7-11}; 2018; Guangzhou, China\string: 1887-1894

\bibitem{Eibl2015}
{Eibl} G, {Engel} D. Influence of Data Granularity on Smart Meter Privacy. {\it
  IEEE Transactions on Smart Grid} 2015\string; 6(2)\string: 930-939.
\newblock \href {\doibase 10.1109/TSG.2014.2376613} {doi:
  10.1109/TSG.2014.2376613}

\bibitem{Cha2019}
{Cha} S, {Hsu} T, {Xiang} Y, {Yeh} K. Privacy Enhancing Technologies in the
  Internet of Things: Perspectives and Challenges. {\it IEEE Internet of Things
  Journal} 2019\string; 6(2)\string: 2159-2187.
\newblock \href {\doibase 10.1109/JIOT.2018.2878658} {doi:
  10.1109/JIOT.2018.2878658}

\bibitem{Parisio2017}
{Parisio} A, {Wiezorek} C, {Kyntäjä} T, {Elo} J, {Strunz} K, {Johansson} KH.
  Cooperative MPC-Based Energy Management for Networked Microgrids. {\it IEEE
  Transactions on Smart Grid} 2017\string; 8(6)\string: 3066-3074.
\newblock \href {\doibase 10.1109/TSG.2017.2726941} {doi:
  10.1109/TSG.2017.2726941}

\bibitem{Chang2017}
{Chang} L, {Wang} X, {Mao} M. Transactive energy scheme based on multi-factor
  evaluation and contract net protocol for distribution network with high
  penetration of DERs. In:  {\it 2017 Chinese Automation Congress (CAC), Oct.
  20-22}; 2017; Jinan, China\string: 7139-7144

\bibitem{Ge2020}
{Ge} Y, {Ye} H, {Loparo} KA. Agent-Based Privacy Preserving Transactive Control
  for Managing Peak Power Consumption. {\it IEEE Transactions on Smart Grid}
  2020\string; 11(6)\string: 4883-4890.
\newblock \href {\doibase 10.1109/TSG.2020.2997314} {doi:
  10.1109/TSG.2020.2997314}

\bibitem{le2018enabling}
Le X, Chen S, Yan Z, et al. Enabling a transactive distribution system via
  real-time distributed optimization. {\it IEEE Transactions on Smart Grid}
  2018\string; 10(5)\string: 4907--4917.

\bibitem{Nizami2020}
{Nizami} MSH, {Hossain} MJ, {Fernandez} E. Multiagent-Based Transactive Energy
  Management Systems for Residential Buildings With Distributed Energy
  Resources. {\it IEEE Transactions on Industrial Informatics} 2020\string;
  16(3)\string: 1836-1847.
\newblock \href {\doibase 10.1109/TII.2019.2932109} {doi:
  10.1109/TII.2019.2932109}

\bibitem{Aitzhan2018}
{Aitzhan} NZ, {Svetinovic} D. Security and Privacy in Decentralized Energy
  Trading Through Multi-Signatures, Blockchain and Anonymous Messaging Streams.
  {\it IEEE Transactions on Dependable and Secure Computing} 2018\string;
  15(5)\string: 840-852.
\newblock \href {\doibase 10.1109/TDSC.2016.2616861} {doi:
  10.1109/TDSC.2016.2616861}

\bibitem{yang2020blockchain}
Yang Q, Wang H. Blockchain-Empowered Socially Optimal Transactive Energy
  System: Framework and Implementation. {\it IEEE Transactions on Industrial
  Informatics} 2021\string; 17(5)\string: 3122--3132.
\newblock \href {\doibase 10.1109/TII.2020.3027577} {doi:
  10.1109/TII.2020.3027577}

\bibitem{yang2021privacy}
Yang Q, Wang H. Privacy-Preserving Transactive Energy Management for IoT-aided
  Smart Homes via Blockchain. {\it IEEE Internet of Things Journal}
  2021\string: 1--13.
\newblock \href {\doibase 10.1109/JIOT.2021.3051323} {doi:
  10.1109/JIOT.2021.3051323}

\bibitem{chen2021distributed}
Chen S, Zhang L, Yan Z, Shen Z. A distributed and robust security-constrained
  economic dispatch algorithm based on blockchain. {\it IEEE Transactions on
  Power Systems} 2021.

\bibitem{chen2021trusted}
Chen S, Shen Z, Zhang L, et al. A trusted energy trading framework by marrying
  blockchain and optimization. {\it Advances in Applied Energy} 2021\string;
  2\string: 100029.

\bibitem{Lu2020}
{Lu} Y, {Lian} J, {Zhu} M. Privacy-Preserving Transactive Energy System. In:
  {\it 2020 American Control Conference (ACC), Jul. 1-3}; 2020; Denver, CO,
  USA\string: 3005-3010

\bibitem{Yang2013}
{Yang} S, {Tan} S, {Xu} J. Consensus Based Approach for Economic Dispatch
  Problem in a Smart Grid. {\it IEEE Transactions on Power Systems}
  2013\string; 28(4)\string: 4416-4426.
\newblock \href {\doibase 10.1109/TPWRS.2013.2271640} {doi:
  10.1109/TPWRS.2013.2271640}

\bibitem{Zhao2017}
{Zhao} C, {He} J, {Cheng} P, {Chen} J. Privacy-preserving consensus-based
  energy management in smart grid. In:  {\it 2017 IEEE Power \& Energy Society
  General Meeting, July 16-20}; 2017; Chicago, Illinois, USA\string: 1-5

\bibitem{Celik2018}
{Celik} B, {Roche} R, {Bouquain} D, {Miraoui} A. Decentralized Neighborhood
  Energy Management With Coordinated Smart Home Energy Sharing. {\it IEEE
  Transactions on Smart Grid} 2018\string; 9(6)\string: 6387-6397.
\newblock \href {\doibase 10.1109/TSG.2017.2710358} {doi:
  10.1109/TSG.2017.2710358}

\bibitem{Wang2019}
{Wang} J, {Zhong} H, {Wu} C, {Du} E, {Xia} Q, {Kang} C. Incentivizing
  distributed energy resource aggregation in energy and capacity markets: An
  energy sharing scheme and mechanism design. {\it Applied Energy} 2019\string;
  252\string: 113471.
\newblock \href {\doibase 10.1016/j.apenergy.2019.113471} {doi:
  10.1016/j.apenergy.2019.113471}

\bibitem{austinenergy}
{\it Residential Electric Rates \& Line Items}. .
\newblock Austinenergy.com accessed Dec. 2, 2020.
\newblock Available:
  https://austinenergy.com/ae/rates/residential-rates/residential-electric-rates-and-line-items/.

\bibitem{Cui2019}
{Cui} S, {Wang} Y, {Xiao} J. Peer-to-Peer Energy Sharing Among Smart Energy
  Buildings by Distributed Transaction. {\it IEEE Transactions on Smart Grid}
  2019\string; 10(6)\string: 6491-6501.
\newblock \href {\doibase 10.1109/TSG.2019.2906059} {doi:
  10.1109/TSG.2019.2906059}

\bibitem{wang2016incentivizing}
Wang H, Huang J. Incentivizing energy trading for interconnected microgrids.
  {\it IEEE Transactions on Smart Grid} 2018\string; 9(4)\string: 2647--2657.
\newblock \href {\doibase 10.1109/TSG.2016.2614988} {doi:
  10.1109/TSG.2016.2614988}

\bibitem{Passey2018}
{Passey} R, {Watt} M, {Bruce} A, {MacGill} I. Who pays, who benefits? The
  financial impacts of solar photovoltaic systems and air-conditioners on
  Australian households. {\it Energy Research \& Social Science} 2018\string;
  39\string: 198-215.
\newblock \href {\doibase 10.1016/j.erss.2017.11.018} {doi:
  10.1016/j.erss.2017.11.018}

\bibitem{lai2014decentralized}
Lai X, Xie L, Xia Q, Zhong H, Kang C. Decentralized multi-area economic
  dispatch via dynamic multiplier-based Lagrangian relaxation. {\it IEEE
  Transactions on Power Systems} 2014\string; 30(6)\string: 3225--3233.

\bibitem{Li2019}
{Li} J, {Ye} Y, {Papadaskalopoulos} D, {Strbac} G. Consensus-Based Coordination
  of Time-Shiftable Flexible Demand. In:  {\it 2019 International Conference on
  Smart Energy Systems and Technologies (SEST), Sep. 9-11}; 2019; Porto,
  Portugal\string: 1-6

\bibitem{Zhang2012}
{Zhang} Z, {Chow} M. Convergence Analysis of the Incremental Cost Consensus
  Algorithm Under Different Communication Network Topologies in a Smart Grid.
  {\it IEEE Transactions on Power Systems} 2012\string; 27(4)\string:
  1761-1768.
\newblock \href {\doibase 10.1109/TPWRS.2012.2188912} {doi:
  10.1109/TPWRS.2012.2188912}

\bibitem{Liu2020}
{Liu} XK, {Yan} J, {Xing} L, {Wen} C. Constrained Consensus-based Iterative
  Algorithm for Economic Dispatch in Power Systems. In:  {\it The 46th Annual
  Conference of the IEEE Industrial Electronics Society, Oct. 19-21}; 2020;
  Singapore\string: 3537-3542

\bibitem{wang2014hybrid}
Wang H, Huang J. Hybrid renewable energy investment in microgrid. In:  {\it
  2014 IEEE International Conference on Smart Grid Communications
  (SmartGridComm)}; 2014\string: 602--607.

\bibitem{wang2015joint}
Wang H, Huang J. Joint investment and operation of microgrid. {\it IEEE
  Transactions on Smart Grid} 2017\string; 8(2)\string: 833--845.
\newblock \href {\doibase 10.1109/TSG.2015.2501818} {doi:
  10.1109/TSG.2015.2501818}

\bibitem{pecan}
{\it Dataport}. .
\newblock Website, Pecan Street Inc. accessed Oct. 1, 2019.
\newblock Available: https://www.pecanstreet.org/dataport/.

\end{thebibliography}

\end{document}